\documentclass{aa}
\usepackage{graphicx}
\begin{document}

\def\einst{{\sl Einstein}}
\def\ros{{\sl ROSAT}}
\def\chan{{\sl Chandra}}
\def\hst{{\sl Hubble Space Telescope}}
\def\obj{\object{1RXS~J115928.5$-$524717}}
\def\lp{\object{LP~944$-$20}}

\title{A new strongly X-ray flaring M9 dwarf in the solar neighborhood\thanks{Based
on observations collected with the ESO 1.54\,m/DFOSC and NTT/EMMI
at the European Southern Observatory, La Silla, Chile (ESO
programmes 69.D-0605 and 69.D-0486).}}

\subtitle{}

   \author{V.\,Hambaryan\inst{1} \and
           A.\,Staude\inst{1} \and
           A.D.\,Schwope\inst{1} \and
           R.-D.\,Scholz\inst{1} \and
           S. Kimeswenger\inst{2} \and
           R.\,Neuh\"auser \inst{3,4}}

   \offprints{V. Hambaryan, vhambaryan@aip.de}

   \institute{Astrophysikalisches Institut Potsdam,
              An der Sternwarte 16, D-14482 Potsdam, Germany
\and
Institut f{\"u}r Astrophysik der Universit{\"a}t Innsbruck, Technikerstr. 25,
A-6020 Innsbruck, Austria
\and
MPI f\"ur extraterrestrische Physik, Gissenbachstrasse 1, D-85740 Garching, Germany
\and
Astrophysikalisches Institut, Universit\"at Jena, Schillerg\"asschen 2-3, 07745 Jena
}

   \date{Received 23 September 2003 / Accepted 7 October 2003}

\abstract {We report on the discovery of a very low mass (VLM)
star in the solar neighborhood, originally identified as an
optical counterpart of a flaring X$-$ray source detected in the
\ros\ All-Sky survey. Optical spectroscopy and infrared photometry
consistently reveal a spectral type of $ M9 \pm 0.5$ and a
distance of $\approx$~11$\pm$~2~pc. The optical counterpart of
\obj\ shows a large proper motion of 1.08$\pm$0.06~$\arcsec$/year.
\obj\ is the fourth object among the VLM stars displaying a huge
X$-$ray flare, reaching the unprecedent value of $L_X/L_{\rm
bol}~\simeq$~0.1.

}

   \maketitle

\section{Introduction}\label{intro}

Imaging sky surveys performed during the last decade
at different wavelengths (optical, near--infrared)
have uncovered a large
population of very low--mass (VLM) stars and brown dwarf
candidates in star forming regions, open clusters
and in the general galactic field.

According to an estimate presented by Reid et al.
(\cite{reid2003}) the current 10~pc sample is only $\simeq $~75\%
complete and the level of incompleteness is somewhat higher in the
southern sky and with the latest type objects.

With spectral types of M7 and later, these objects, sometimes
called ``ultracool M dwarfs'', are so faint optically that even
nearby ones eluded searches based on the optical, near--infrared
and high proper motion sky surveys, and, therefore a detection of
a new VLM star in the solar vicinity is of vital importance in the
determination of fundamental parameters such as the luminosity
function, the mass function, the kinematics of those stars and
other properties.

The strength of H${\alpha}$-emission increases as cooler,
lower-mass stars are considered (Hawley et al. \cite{haw96}; Gizis
et al. \cite{gizis00}). The emission is up to 100\% at spectral
types M7-M9. The observed decline in H${\alpha}$-emission for old
($>$1\,Gyr) L-type dwarfs in the field suggests that the activity
borderline is close to  objects of substellar masses (Gizis et
al. \cite{gizis00}; Basri \cite{basri01}; Mohanty \& Basri
\cite{mohanty}).

Analyzing X$-$ray properties of VLM stars based on \ros\
observations, Fleming et al.~(\cite{Fleming93}) showed that there
is no apparent decrease in X$-$ray flux (hence coronal heating
efficency) down to spectral type M5, as was suggested by Mullan
(\cite{mullan}) analyzing \einst\ data.

Flares as another property  of chromospheric and coronal activity
represent a phenomenon which is important in ultracool dwarfs,
because at these low temperatures (and masses) these objects are
fully convective, and variability is often attributed to
rotational modulation of star spots produced by magnetic activity.
In solar-type stars it is believed to be due to the so-called
$\alpha \Omega$ dynamo. This mechanism no longer operates in VLM
stars and brown dwarfs, but as these objects are fully convective,
a turbulent dynamo could come into operation (see Chabrier \&
Baraffe \cite{chabrier00a} and references therein). This means
that the change in interior structure is expected to result in a
change of the field sustaining dynamo, and, therefore X$-$ray
emission properties of late-type stars.

Flares were reported for a number of VLM stars, observed in
ultraviolet, X$-$ray, and radio frequencies, as well as in optical
spectra or via  photometry. Bursts of X$-$ray emission 
interpreted as coronal flares were
detected from the long-known M8V dwarf \object{vB 10} (Fleming et
al. \cite{vb10}), the M9V dwarf \object{LHS~2065} (Schmitt \&
Liefke \cite{lhs2065}) by \ros\ and the M9 brown dwarf
\object{LP~944$-$20} (Rutledge et al. \cite{lp944}) by \chan\
observations.

\object{vB 10} showed a far-ultraviolet flare observed by Linsky
et al. (\cite{linsky}) using the GHRS on the \hst . Radio emission
has been observed from \lp\ and some other ultracool dwarfs 
(Berger et al. \cite{berger01}, Berger \cite{berger02}). 
From the intensity of
the continuous emission and flaring at 8.5\,GHz detected in these
objects, Berger (\cite{berger02}) concluded that the emission
mechanism is most likely synchrotron radiation and not of thermal
origin.  This indicates the presence of magnetic fields and electron
column densities similar to those inferred for earlier-type
flaring M dwarfs.

H${\alpha}$-flares were observed on several occasions, e.g. in the
M8.5 \object{APMPM~J2354$-$3316CM9.5} (Scholz et al.
\cite{scholz2003a}) and M9.5 dwarfs
\object{2MASSW~J0149090$+$295613} (Liebert et al.
\cite{liebert99}), \object{BRI~0021$-$214} (Reid et al.
\cite{reid99})  and even in the cooler L5-dwarf
\object{2MASS~J01443536$-$0716142} (Liebert et al.
\cite{liebert03}).

Thus, it has become apparent that the level of quiescent
chromospheric and coronal activity as measured by the H${\alpha}$
emission line and X$-$ray fluxes relative to the total bolometric
flux generally declines with spectral type later than M7.

However, {\emph no} clear change in flare activity diagnostics is
found in VLM stars (Fleming et al. \cite{Fleming95}; Mokler \&
Stelzer \cite{mokler}; Mart{\'\i}n \& Bouy \cite{martin}).

We performed a systematic search for X$-$ray variability in the
\ros\ All-Sky survey data (Hambaryan et al. \cite{skymi}). Our
technique is customized for the detection of flare-like
variability. Here we report on the identification of the optical
counterpart of the strong flaring X$-$ray source \obj\ as a new
ultracool dwarf star in the solar neighborhood by optical
spectroscopy and astrometry.

\section{Observational data and reduction}\label{data}

\subsection{X$-$ray data}\label{xray}

During the \ros\ All--Sky Survey (RASS) the whole sky was observed
in an unbiased fashion. This, therefore, is well suited to study
of variability in different types of celestial objects
in the X$-$ray range (0.1$-$2.4\,\,keV).

Using a Bayesian change point detection method (see
Sect.~\ref{xray_fl}), we analyzed the RASS Bright Source Catalog
(BSC) for variability. The method works on unbinned data, i.e. the
photon arrival times of RASS event tables. We found that out of
18811 BSC sources 642 show significant variations (Hambaryan et
al. \cite{skymi}).

\begin{figure*}
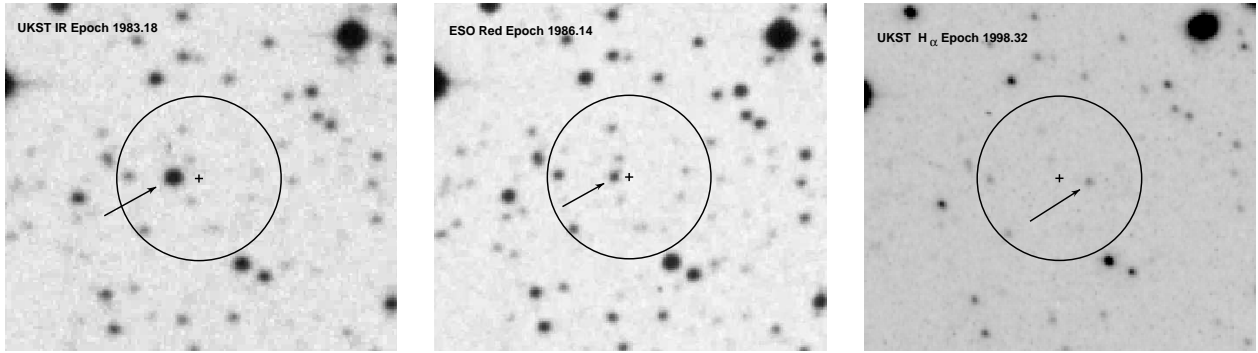

\centering \hbox{
\includegraphics[bbllx=72, bblly=220,bburx=524, bbury=622,width=5.2cm,clip]{0378.f1a}
\hspace{0.3cm}
\includegraphics[bbllx=72,bblly=220,bburx=524,bbury=622,width=5.2cm,clip]{0378.f1b}
\hspace{0.3cm}
\includegraphics[bbllx=72,bblly=220,bburx=524,bbury=622,width=5.2cm,clip]{0378.f1c}
} \caption{$2{\arcmin} \times 2{\arcmin}$ SuperCOSOMOS Sky Survey I,
R and H$_{\alpha}$ images of the field of RASS X$-$ray source
\object{1RXS~J115928.5$-$524717} in 1983 (left), 1992(middle) and
1998 (right) showing the relative proper motion between three
epochs. \obj\ RASS position is depicted by a cross and the 90\%
confidence error circle is shown with a radius of 25${\arcsec}$
(based on the \ros\ BSC statistics; Voges et al.~\cite{bsc99}).
The arrow marks the most probable optical counterpart with its
high proper motion and red color.} \label{charts}
\end{figure*}

In the framework of this project the X$-$ray source \obj\ was
identified as  variable\footnote{\scriptsize This source was
included in the search list of X$-$ray afterglows from gamma--ray
bursts by Greiner et al. (\cite{jochen}). It is also included in the
catalogue of variable sources of RASS as a flaring object
(Fuhrmeister \& Schmitt \cite{fuhr}).}. Our identification
procedure of optical counterparts of RASS sources typically
consists of several steps and first includes the automatic
extraction of DSS images and a visual search for possible
counterparts in an error circle with radius 25$\arcsec$. In this
particular case, as one sees in Fig.~\ref{charts}, there is no
unique optical counterpart of \obj\ and our dedicated follow$-$up
optical spectroscopy in such a case starts with the brightest
object in the X$-$ray error circle. However, while inspecting
finding charts obtained at different epochs, we found that the
brightest object in the DSS II R-band image (epoch  1991.12, the
closest one to the RASS observation, cf.~Table~2) has changed its
position significantly. Further investigation showed that it is a
high proper motion star and also bright on an H${\alpha}$ image
taken by the UKST (epoch 1998.32).

Other fainter objects in the X$-$ray error circle cannot be
completely removed from further consideration as the
counterpart of the \object{1RXS~J115928.5$-$524717}. Nevertheless,
we concentrated in our follow$-$up study on the brightest one.

The RASS was performed in such a way that the sky was scanned
along great circles with the \ros\ X$-$ray telescope; the scanning
period was equal to the orbital period, i.e. $\approx$~96 minutes.
A given source was scanned for at least 2 days and during a single
scan, it remained in the field of view for typically 10$-$30
seconds.

The X$-$ray source \obj\ is included in the RASS Source Catalogue
(1RXS) with a count rate of $0.10060 \pm 0.02097$~cts~s$^{-1}$ and
was observed during January 9$-$11, 1991 (ROR number 932622).

The arrival times of photons registered in the RASS during
this observation of \obj\  were extracted
using the position and extraction radius available in the
BSC (Voges et al.~\cite{bsc99}). The Good Time Intervals 
were determined in a manner described by
Belloni et al. (\cite{bellonietal}), in which the whole extraction
circle ($r = 300\arcsec$) was completely within the field of view
of the detector.

\subsubsection{Optical data}

Low-resolution spectroscopy and photometry were performed with
DFOSC at the 1.54\,m Danish telescope at ESO on July 14 and 17,
2002. The nights were not exactly photometric but sufficient for
the spectroscopy. We obtained three spectra of \obj\ with the DFOSC
grism \#5 and grism \#15.

Standard calibration exposures were taken for bias subtraction,
flat fielding and flux calibration with spectrophotometric
standard stars. Wavelength calibration was done with arc-lamp
spectra and additionally inspected with the night-sky lines.

We also obtained two medium-resolution spectra of the likely
counterpart of \obj\ on 4 August, 2002 at the 3.5\,m-New
Technology Telescope (NTT) of European Southern Observatory (ESO)
at La Silla with the ESO Multi Mode Instrument (EMMI) in the red
medium dispersion mode. Grating \#6 was used, providing a FWHM
spectral resolution of $4-5$\,\AA\ (with the 1.0${\arcsec}$ slit)
and covered a spectral range of $6000-9100$\,\AA.

All reductions were performed with MIDAS. The CCD 44-82 currently
in use with DFOSC is a thinned chip, so there is strong fringing
seen on the red side of 7000 \AA, and it is also visible in
Gunn-$i$ images. For the spectra, the fringing could be
satisfactory removed with the flat-field below 8000 \AA. For the
$i$-band images the dome flat-field correction removed the fringes
completely.

\section{Data analysis and Results}\label{results}

\subsection{X$-$ray flare}\label{xray_fl}

\subsubsection{Light curve}

In order to study the time behavior of \obj\ we performed a timing
analysis of the dataset using a Bayesian change point detection
approach developed by Scargle (\cite{scar1,scar2}). It is very
well suited for a statistical examination when the arrival times
of individual X$-$ray photons are registered (see Hambaryan
et\,al. \cite{arajin}; Schwope et\,al. \cite{erkrord}).

The essence of the method is that it subdivides a given data set
into intervals with a piecewise constant X$-$ray count rate
according to Poissonian statistics. The  application of 
this procedure to the RASS-detected photons
finds three change points, i.e.~the data set was decomposed into
four observational segments with no variation within them and
highly significant variations of count rates between them.

\begin{figure}[t]
\centering
\includegraphics[width=8.8cm,clip]{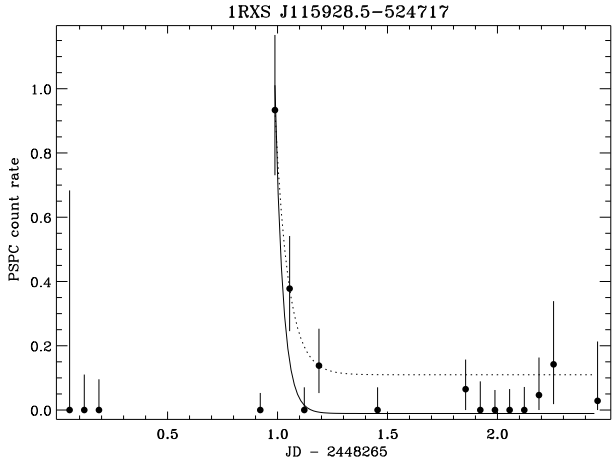}

\caption{The X$-$ray light curve is displayed for the \ros\
All-Sky survey observation of \obj\ . Solid and dashed lines 
represent exponential decay with and without the data point
corresponding to OBI \#7, with vignetting$-$corrected peak count
rate = 0.93.} \label{lcu}
\end{figure}

Fig.~\ref{lcu} and Table~\ref{tbl1} show the RASS X$-$ray
lightcurve of \obj\ in the form of average count-rate for each
observational interval.
In Fig.~\ref{lcu} the PSPC count rate is plotted vs. time in
Julian days. The most dominant light curve feature is a giant
X$-$ray flare that occurred between JD 2448266.0 and JD 2448266.3.
Owing to the data gaps and sparse statistics it is not possible to
determine the flare onset time.

In total (source+background), 52 photons were registered during the
RASS observations. As a background level we used a value given
in the \ros\ BSC (Voges et al. \cite{bsc99}) for this observation,
estimated for the field of view (vignetting corrected
0.00087 counts/sec/arcmin$^2$).
Thus, from these, 52 photons 32$-$3.68 $\simeq$~28 are attributed
to the flare of \obj. They were recorded in the two subsequent
observational intervals OBI \# 5 and 6 (cf.~Table~\ref{tbl1}). In
OBI \#7 just one photon was registered, while during the next one
(OBI \# 8) out of 5 registered photons 3 are attributed to the
source. In order to estimate the intrinsic source count rate and
confidence limits for each observational interval we used a
Bayesian approach for low numbers of counts developed by Kraft et
al.~(\cite{kraft}). Instead of the widely used approach of simply
subtracting the mean number of background counts from the observed
number of counts it correctly takes into account Poissonian
fluctuations in the number of counts.

In the last two columns of Table~\ref{tbl1} a mode of the probability
distribution function of vignetting$-$corrected intrinsic source
count rates and credible regions of the source count rates are
given. In Fig.~\ref{lcu} these Bayesian credible regions
(posterior bubble 0.6827) for each estimated value of the source
count rate are plotted.

An estimate of the quiescent X$-$ray count rate is possible after
exclusion of the flare OBIs. Excluding OBIs \#5 and 6 and
application of the above procedure to all remaining photons gives
an upper limit to the count rate of 0.015\,s$^{-1}$, excluding OBI
\#8 as well gives an upper limit of 0.011\,s$^{-1}$.
It should be noted that the source was clearly detected during
three observational intervals (OBI \# 5,6 and 8).

\begin{table}
\caption[]{\ros\  All-Sky survey  X$-$ray observations of
\obj.} \label{tbl1}
\begin{tabular}{cllcrccc}
\hline
$\!\!\!$OBI$\!\!\!$& ~~~JD        & $\!\!$Expo$\!\!\!$ & $\!\!\!\!$cts$\!\!\!$ & Back-& $\!\!$Source$\!\!$& $\!\!\!\!$Credible$\!\!\!\!$ \\
     & +2448266& ~(s)  & $\!\!\!\!\!$(s+b)$\!\!\!$ & $\!\!\!$ground$\!\!\!\!$ &  & regions  \\
\hline
$\!\!$~\,1 & $\!$0.05456019 &  ~9.00   & ~\,1    & 0.613   & 0.054  & 0.000-0.269$\!\!\!\!$\\
$\!\!$~\,2 & $\!$0.12127315 &  13.00   & ~\,0    & 0.885   & 0.000  & 0.000-0.110$\!\!\!\!$\\
$\!\!$~\,3 & $\!$0.18798611 &  15.00   & ~\,0    & 1.021   & 0.000  & 0.000-0.096$\!\!\!\!$\\
$\!\!$~\,4 & $\!$0.92155093 &  27.00   & ~\,0    & 1.838   & 0.000  & 0.000-0.053$\!\!\!\!$\\
$\!\!$~\,5 & $\!$0.98826389 &  27.00   & 22    & 1.838   & 0.933  & 0.731-1.167$\!\!\!\!$\\
$\!\!$~\,6 & $\!$1.05497685 &  27.00   & 10    & 1.838   & 0.378  & 0.246-0.542$\!\!\!\!$\\
$\!\!$~\,7 & $\!$1.12168981 &  28.00   & ~\,1    & 1.906   & 0.000  & 0.000-0.071$\!\!\!\!$\\
$\!\!$~\,8 & $\!$1.18840278 &  28.00   & ~\,5    & 1.906   & 0.138  & 0.053-0.253$\!\!\!\!$\\
$\!\!$~\,9 & $\!$1.45490741 &  28.00   & ~\,1    & 1.906   & 0.000  & 0.000-0.071$\!\!\!\!$\\
$\!\!$10 & $\!$1.85520833 &  25.00   & ~\,3    & 1.702   & 0.065  & 0.000-0.157$\!\!\!\!$\\
$\!\!$11 & $\!$1.92192130 &  23.00   & ~\,1    & 1.566   & 0.000  & 0.000-0.089$\!\!\!\!$\\
$\!\!$12 & $\!$1.98863426 &  23.00   & ~\,0    & 1.566   & 0.000  & 0.000-0.062$\!\!\!\!$\\
$\!\!$13 & $\!$2.05534722 &  22.00   & ~\,0    & 1.498   & 0.000  & 0.000-0.065$\!\!\!\!$\\
$\!\!$14 & $\!$2.12206019 &  20.00   & ~\,0    & 1.362   & 0.000  & 0.000-0.072$\!\!\!\!$\\
$\!\!$15 & $\!$2.18877315 &  19.00   & ~\,2    & 1.294   & 0.046  & 0.000-0.163$\!\!\!\!$\\
$\!\!$16 & $\!$2.25520833 &  11.00   & ~\,2    & 0.749   & 0.142  & 0.018-0.339$\!\!\!\!$\\
$\!\!$17 & $\!$2.45530093 &  11.00   & ~\,1    & 0.749   & 0.029  & 0.000-0.213$\!\!\!\!$\\
\hline
\end{tabular}
\end{table}

Assuming an exponential flare decay we determined the flare decay
e-folding time for two cases, either including or excluding OBI
\#7. This can be justified by assuming that the flaring structure
is hidden behind the star due to rotation. In the first case the
e-folding time is $\tau_{LC}$ = 0.04 days, in the second case it
is 0.07 days.

\subsubsection{Spectral analysis}

Despite the observed small number of counts (Source+Background=52
and Background=20) we performed a crude spectral analysis, binning
the data with a constant Signal/Noise ratio of 2, i.e. into bins,
which contained between 3 and 7 (S+B) and between 0 and 2 (B)
counts, respectively, in the energy range 0.11$-$2.46~keV. Note
that no photons were registered above 1.25~keV in the S+B region.
We fit the resulting spectrum with an assumed Raymond-Smith plasma
model with negligible absorption, implemented in XSPEC version
11.1 (Arnaud \cite{arnaud}). Indeed, applying $N_H \approx 0.07 {\rm
\,\,cm^{-3}}$ from Paresce (\cite{P84}) and using a distance of
11.1 pc (see \ref{absmag}) we obtain N$_H = {\rm 2.4}\times
10^{-18} {\rm \,\,cm^{-2}}$ . With this fixed interstellar
absorption parameter the best fit according to the C statistics
gives ${\rm kT} = 0.20_{-0.12}^{+0.10} {\rm \,\,keV}$ (90\%
confidence) and a time$-$averaged unabsorbed flux of $8.1 \times
10^{-13} {\rm \,\,erg\,\, cm^{-2}\,\,s^{-1}}$.

An application to the data of OBI~\#5 and \#6 separately gives
${\rm kT} = 0.28_{-0.11}^{+0.20} {\rm \,\,keV}$ and unabsorbed
flux of ${\rm f_X} = 6.2 \times 10^{-12} {\rm \,\,erg \,\,cm^{-2}
\,\, s^{-1}}$ in the energy range 0.1-2.4~{\rm \,\,keV} for OBI
\#5 and ${\rm kT} = 0.23_{-0.20}^{+0.27} {\rm \,\,keV}$ and ${\rm
f_X} = 2.9\times 10^{-12} {\rm \,\,erg\,\,cm^{-2} \,\, s^{-1}}$
for OBI \#6, respectively.

\subsection{Optical spectroscopy}
\subsubsection{Spectral type and absolute magnitude}\label{absmag}

The spectrum of the proper motion object taken by us with DFOSC is
shown in Fig.~\ref{fig_2spec}. For comparison we show the
spectra of the late M dwarfs vB\,8 (spectral type M7) and \lp\
(spectral type M9.5) taken with the same instrumental set up in
the same observing run. The likely counterpart of \obj\ shows
similar spectral features as the two template stars.

\begin{figure*}
\centering
\hbox{
\includegraphics[width=6.0cm,angle=90,clip]{0378.f3a}
\hspace{0.1cm}
\includegraphics[width=6.2cm,angle=90,clip]{0378.f3b}
} \caption{ESO 1.54-m/DFOSC spectra (left panel) of \obj\ compared
with those of LP\,944-\,20, a known M9.5 brown dwarf (Tinney
\cite{tinney96,tinney98}; Reid et al. \cite{rhg95}), and the M7.0
dwarf vB8 (Gizis et al. \cite{gizis00}). An arbitrary constant has
been used to separate the spectra. The location of features
typical of late-type stars are labelled, including metal oxide and
hydride absorption bands.  A fragment of the \obj\ ESO NTT
EMMI+Grism\#6 spectrum (right panel) with H$\alpha$ emission and
lithium absorption location is provided. } \label{fig_2spec}
\end{figure*}

\begin{figure*}
  \resizebox{12cm}{!}{\includegraphics[angle=90]{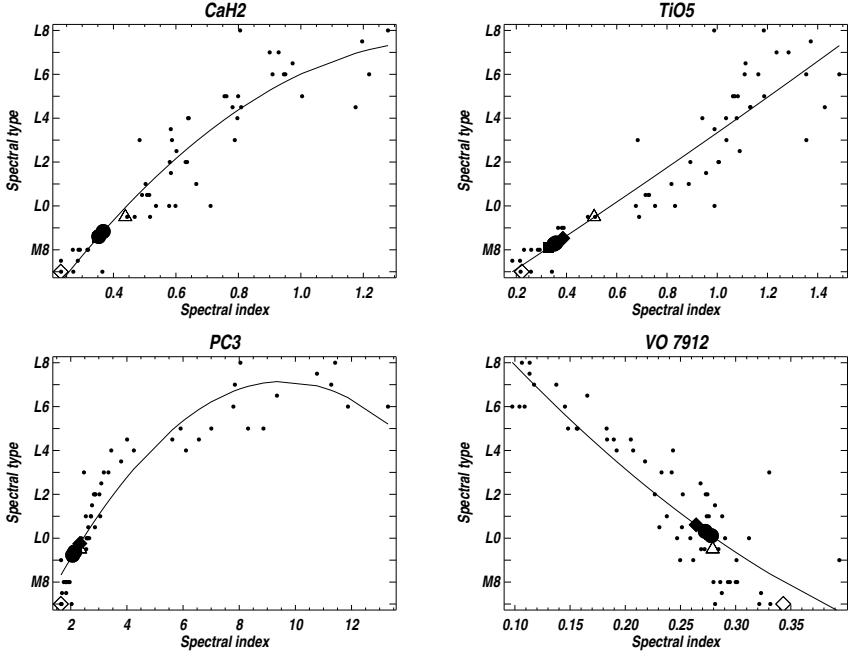}}
  \hfill
  \parbox[b]{55mm}{
   \caption{Relationships between spectral type and the spectral
index for 4 different spectral indices, showing M7-L8 dwarfs
(small, filled circles). The lines are second-order polynomial,
least-squares fits. The open diamond corresponds to the star vB8,
the open triangle to the brown dwarf LP944$-$20. 
The spectral type estimates of \obj \ with filled
large symbols are depicted using 1.54m Danish DFOSC (diamond and triangle) 
and NTT EMMI (circle) spectra.} \label{spi}
}
\end{figure*}

To assign a spectral type to our object we used two
different approaches based on the strengths of spectral features
and the slope of the spectrum. First we compiled a set of spectral
standards using 74 late type dwarfs (from M0 to L8)\footnote{
\scriptsize available at http://www.physics.upenn.edu/$\sim$inr
and\protect\newline
http://www.astro.washington.edu/covey/research/research.html} and
then employed spectral classification methods. In the first
method, spectral indices of individual features (in total 16) were
measured and compared with a standard sequence, following Cruz \&
Reid (\cite{cruz_reid}) and Hawley et al.~(\cite{haw02}). We
computed their spectral indices to determine the
relationships between spectral type and individual spectral
features, suggested by the same authors. In Fig.~\ref{spi} some of
these relationships are shown.

The second method, described by Henry et al.~(\cite{h02}), tries
to match the overall shape of the observed spectrum to template
spectra (see also Hawley et al. \cite{haw02}).

The method works very well for the known late-type M dwarfs \lp\
and vB\,8 also observed by us. We could reproduce the published
spectral subclass with an uncertainty of half a subclass (see
Fig.~\ref{spi}).

For each of these relations one gets an independent estimate of
the spectral class. We take the spectral type to be the average of
those estimates rounded to the nearest half spectral type. The
resulting spectral type of \obj\  determined consistently by both
methods is M9 $\pm$0.5.

We identified \obj\ with the Two Micron All Sky Survey source
\object{2MASS~11592743$-$5247188} with magnitudes
$J=11.430\pm0.023, H=10.763 \pm 0.021$ and $ K_s = 10.322 \pm
0.017$. Following Gizis et al.~(\cite{gizis00}) we estimate $M_K$
from the $J-K_s$ color using the relation
\begin{displaymath}
M_K = 7.593 + 2.25 \times (J-K_s),
\end{displaymath}
with a scatter of $\sigma = 0.36$ magnitudes. The relation is
valid for dwarfs in the color range $1.0 \le$ J-K$_S \le 1.6$
corresponding typically to dwarfs later than M7. The relation
gives  $M_K = 10.08 \pm 0.36$, the implied distance to \obj\ is
$d_{M_K} = 11.1 \pm 2 {\rm pc}$.

If we instead derive the absolute magnitude of the likely
counterpart of \obj\ via the spectral type,
\begin{displaymath}
M_J = 8.38 + 0.341 \times {\rm Sp.type},
\end{displaymath}
(Dahn et al.~\cite{dahn}), we arrive at very similar results.
Indeed, for spectral types M8.5, M9 and M9.5 we get
$d_{M_J}^{M8.5} = 9.6-12.0 {\rm pc}$, $d_{M_J}^{M9} = 8.8-11.1 {\rm pc}$
and $d_{M_J}^{M9.5} = 8.2-10.3 {\rm pc}$, correspondingly.

\subsection{Proper motion}

While the large proper motion of the object was already discovered
during the visual inspection of Digitized Sky Survey data at two
epochs only, an accurate proper motion determination was done on
the basis of a larger set of measurements with different epochs
available in SuperCOSMOS Sky Survey (SSS) data (Hambly et
al.~\cite{hambly01a}, \cite{hambly01b}, \cite{hambly01c}). With a
location in the overlap region between two UK Schmidt telescope
(UKST) survey fields we were lucky to identify the object in the
SSS catalogues at eight different epochs with three different
passbands ($2 \times B_J$, $4 \times R$ and $2 \times I$). In
addition to the SSS data, two further measurements from the
SuperCOSMOS H-alpha survey (SHS) (Parker \&
Phillips~\cite{parker98}) were available (see
Table~\ref{sssdata}).

The position of the object in the recently completed all sky data
release of the Two Micron All Sky Survey (2MASS, Cutri et al.
\cite{cutri03}) and in the second release of DENIS (DEep
Near-Infrared Survey, Epchtein et al. \cite{denis}) now available
at http://vizier.u-strasbg.fr/viz-bin/Cat?B/denis was included in
the proper motion solution. The 2MASS $JHK_S$ and DENIS $IJK$
photometry is also listed in Table~\ref{sssdata}.

The most recent epoch astrometry was obtained from DFOSC
acquisition images in the $R$ and $I$ bands. The astrometric
calibration was done using the ESO Skycat tool with 16 faint SSS
reference stars located around the target, which did not show
significant proper motions and which were not affected by image
crowding in the SSS data. Two independent astrometric calibrations
for the DFOSC $R$ and $I$ image yielded very similar results. The
average result was used as one additional epoch in the proper
motion solution.

\begin{table}
\caption{Positions and photometry of
\object{1RXS~J115928.5-524717} measured at different epochs}
\label{sssdata}
\begin{tabular}{llcrl}
\hline
~~$\alpha$(J2000)& ~~$\delta$(J2000)& epoch    & photometry & $\!\!\!\!$note \\
\hline
11 59 30.178 & -52 47 15.85 & 1976.26 & $B_J$=21.00 & 1 \\
11 59 30.147 & -52 47 17.04 & 1977.41 & $B_J$=20.75 & 1,$*$ \\
11 59 29.665 & -52 47 16.28 & 1980.23 & $I$=14.49   & 1 \\
11 59 29.340 & -52 47 16.65 & 1983.18 & $I$=13.76   & 1 \\
11 59 29.089 & -52 47 16.92 & 1985.19 & $RE$=17.70  & 2 \\
11 59 28.996 & -52 47 16.55 & 1986.14 & $RE$=17.44  & 2,$*$ \\
11 59 28.425 & -52 47 17.45 & 1991.12 & $OR$=17.66  & 1,$*$ \\
11 59 28.268 & -52 47 17.98 & 1992.24 & $OR$=18.07  & 1 \\
11 59 27.531 & -52 47 18.64 & 1998.32 & $RHa$=17.30  & 3 \\
11 59 27.515 & -52 47 18.56 & 1998.39 & $SR2$=17.96  & 3 \\
11 59 27.43  & -52 47 18.8  & 1999.36 & $J$=11.430 & 4 \\
             &              &         & $H$=10.763 & 4 \\
             &              &         & $K_S$=10.322& 4 \\
11 59 27.32  & -52 47 19.0  & 2000.78 & $I=$14.499 & 5,$*$ \\
             &              &         & $J=$11.398 & 5 \\
             &              &         & $K=$10.343 & 5 \\
11 59 27.047 & -52 47 19.32 & 2002.55 & $R=$17.08  & 6 \\
\hline
\end{tabular}
\smallskip

\noindent
Notes: \\
1 -- SSS measurements of UKST plates. \\
2 -- SSS measurements of ESO Schmidt plates. \\
3 -- SHS film scans. \\
4 -- 2MASS all-sky data release. \\
5 -- DENIS second data release. \\
6 -- Position from two DFOSC acquisition images ($R$ and $I$). \\
$*$ -- Measurements not used in final proper motion solution, due to
       overlap with a background object. Photometry is also affected. \\
\end{table}

Since our object (\object{1RXS~J115928.5-524717}) is moving across
a crowded field of background stars, some of the SSS measurements
are affected by overlapping images, as can also be noted in
the SSS image parameters. There also seems to be a large error
($\approx$1\arcsec) in the $\alpha$ coordinate of the DENIS data.
Therefore, we excluded three of the SSS positions and the DENIS
position given in Table~\ref{sssdata} from the final proper motion
solution, yielding a significant improvement in the proper motion
accuracy (see Table~\ref{pmsol}).

\begin{table}
\caption{Proper motion solutions} \label{pmsol} {
\begin{tabular}{lll}
\hline
positions used in solution & ~~~$\mu_{\alpha}\cos{\delta}$ & ~~~~~~$\mu_{\delta}$ \\
                           & \multispan{2}{\hfil [mas/yr] \hfil} \\
\hline
  13 (all available)                 & $-$1083.9$\pm$11.6$\!\!\!\!\!$& $-$119.0$\pm$11.3 \\
  12 (excl. epoch 1977)              & $-$1069.8$\pm$7.8  & $-$133.8$\pm$6.3 \\
  11 (excl.  1977, 1986)             & $-$1069.5$\pm$8.4  & $-$130.9$\pm$3.9 \\
  10 (excl. 1977, 1986, 1991)        & $-$1069.4$\pm$8.5  & $-$130.7$\pm$2.5 \\
  ~9 (excl. 3 SSS and DENIS)$^{\#}$ $\!\!\!\!\!\!\!$ & $-$1076.6$\pm$5.1  & $-$130.7$\pm$2.9 \\
\hline
\end{tabular}
\smallskip

\noindent Note: $^{\#}$ -- finally adopted proper motion
solution.}
\end{table}

\begin{displaymath}
\,\,\mu_{\alpha}\cos{\delta} = -1076.6 \pm 5.1  \,\,[{\rm mas/yr}]
\end{displaymath}
\begin{displaymath}
\,\,\mu_{\delta}\,\,\phantom{\cos{\delta}} = \,\,\,-130.7 \pm 2.9
\,\,[{\rm mas/yr}]
\end{displaymath}

\section{Discussion and summary}\label{discussion}

The object \obj\ was detected in the RASS due to the presence of a
huge X$-$ray flare. This was uncovered by our time variability
analysis of all RASS-BSC sources. The quiescence X$-$ray flux is
consistent with the background level.

Our follow-up optical spectroscopy and proper motion study allowed
us to determine the spectral type, M$9\pm0.5$, and the absolute
magnitude $M_K= 10.1\pm 0.4$ of the most likely optical
counterpart of the flaring X$-$ray source. It turned out that this
object is a nearby, $d \approx 11\pm2$\,pc, high proper motion,
$\mu =$ 1.08$\pm$0.06$\arcsec$/yr, previously unknown member of
the solar neighborhood.

In order to compare the X$-$ray properties of \obj\ with other
ultracool stars we estimated $\log{(L_{X}/L_{bol})}$. From a
spectral fit (Sect.~\ref{xray_fl}) we estimated time averaged and
flare peak X$-$ray luminosities of $L_{X}^{mean} \approx (0.7-1.9)
\times 10^{28} {\rm \,\,erg\,\,s^{-1}}$ and $L_{X}^{peak} \approx
(1.0-1.6) \times 10^{29} {\rm \,\,erg\,\,s^{-1}}$, using
unabsorbed fluxes in the passband $0.1-2.4$\,\,keV. The luminosity
uncertainties were determined by a combination of uncertainties of
the distance and of the parameters of the spectral fit.

For an estimate of an upper limit of the X$-$ray luminosity of
\obj\ in its quiescent state we converted from upper limit count
rates (0.011 and 0.015) to fluxes by means of an energy conversion
factor $ECF \approx 7.0\times 10^{-12} {\rm \,\,erg\,\,cts^{-1}
cm^{-2}}$ which was computed by us in XSPEC assuming a
Raymond-Smith plasma (model {\emph raymond \/}) with temperature
${\rm kT = 0.25 \,\,keV}$ and absorption column density N$_H =
{\rm 2.4} \times 10^{-18} {\rm cm^{-2}}$ (see also H{\"u}nsch et
al.~\cite{huensch99}, Fleming et al.~\cite{fleming03}). Using a
distance of $11.1 \pm 2.0$\,pc we arrived at an upper limit value
of $L_{X}^{qui.} \approx (1.6-2.0) \times 10^{27} {\rm \,\,erg
\,\,s^{-1}}$ for the quiescent X$-$ray emission of \obj.

Dahn et al.~(\cite{dahn}) list bolometric magnitudes for two M8.5,
five M9 and three M9.5 dwarfs in the immediate solar
neighborhood\footnote{\scriptsize $M_{bol}=13.39$ for
\object{CTI~0126+28}; $M_{bol}=13.77$ for \object{T513-46546};
$M_{bol}=13.33$ for \object{BRI~1222-12}; $M_{bol}=13.44$ for
\object{T868-110639}; $M_{bol}=13.47$ for \object{LHS~2065};
$M_{bol}=13.65$ for \object{LHS~2924}; $M_{bol}=14.22$ for
\object{LP~944-20}; $M_{bol}=13.33$ for \object{BRI~0021-02};
$M_{bol}=13.40$ for \object{2M~0149+29}; and $M_{bol}=13.81$ for
\object{PC~0025+04}.}. An average gives
$M_{bol}=13.58\pm0.28$ magnitudes.

This leads to the  $L_{bol} \approx (0.9-1.5) \times 10^{30}
\rm{erg\,\,s^{-1}}$. Therefore, we estimate the ratios
$\log{(L_{X}/L_{bol})} \approx -1.1 $ at the flare peak and an
upper limit of $\log{(L_{X}/L_{bol})} \approx -3.0 $ in the
quiescent state which can be compared with measurements of other
late-type dwarfs.

\begin{figure*}
  \resizebox{12cm}{!}{\includegraphics{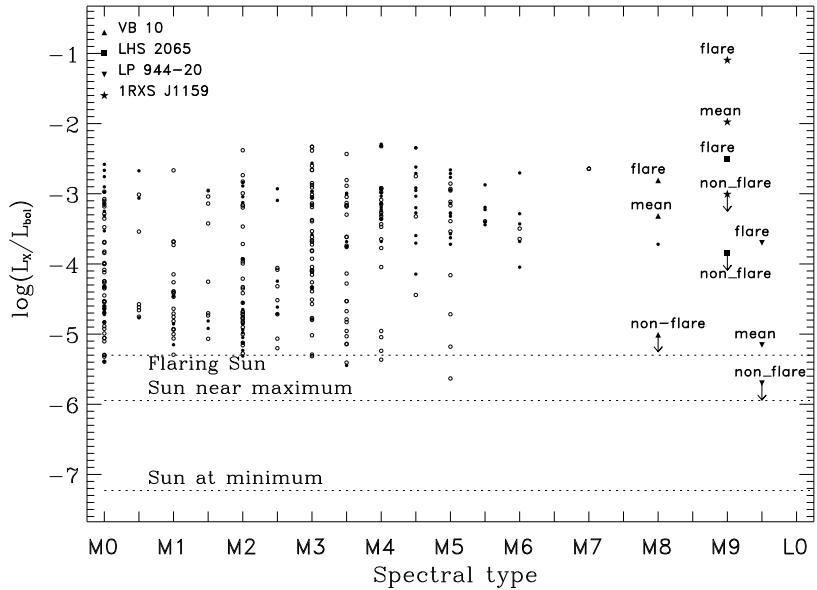}}
  \hfill
  \parbox[b]{55mm}{
\caption{ The X$-$ray
emission level of late-type dwarfs in the solar neighborhood (see,
for details text). Filled symbols indicate H$\alpha$ emission line
stars, open symbols indicate stars without reported line emission
(see H\"unsch et al.~\cite{huensch99}). For the ultracool dwarfs
(\object{vB10}, \object{LHS~2065}, \object{LP~944-20} and \obj)
the quiescent, flare and time averaged X$-$ray luminosity levels
are indicated. } \label{fig1_im}
}
\end{figure*}

In Fig.~\ref{fig1_im} we display the dependence of X$-$ray
activity on spectral type for M dwarfs, using data from the
RASS catalogue of nearby stars (H\"unsch et al. \cite{huensch99}).
To derive bolometric magnitudes we used the relation
\begin{displaymath}
M_{bol} = 7.315 \times (B-V) - 2.238,
\end{displaymath}
(Mullan  \cite{mullan}) which has a typical scatter of 0.36 mag.

We also include in this diagram the only four known examples of
VLM stars with detected X$-$ray emission in the solar
neighborhood. All these stars have quiescent X$-$ray fluxes which
are compatible with zero. They all show pronounced X$-$ray flares,
the newly discovered object \obj\ being the most luminous among
them.

The \ros\  surveys of both field and cluster stars show a maximum
value of about $L_{X}/L_{bol}\simeq 10^{-3}$. Since this ratio may
be regarded as an efficiency measure of the coronal heating
processes and of the activity related phenomena, the maximum value
indicates a still not well understood saturation
phenomenon.

While the final word on the quiescent X$-$ray emission down to the
bottom of the main sequence at spectral type M8/9 is not known, 
(see also the discussions in Fleming et al.
\cite{Fleming95,fleming03}; Mokler \& Stelzer \cite{mokler};
Mart\'\i n \& Bouy \cite{martin}) the flare observations of the
very late M dwarfs suggest that their X$-$ray luminosity is
independent of the spectral subclass.

Any nearby ultracool dwarf, like this new one, is a promising
target for direct imaging searches for sub-stellar companions,
because the primary is nearby (i.e. high physical separation) and
faint (i.e. no overshining companions)\footnote{\scriptsize See
e.g. Neuh\"auser et al (\cite{ralph_denis}) on
\object{DENIS-P~J104814.7$-$395606} and Scholz et al.
(\cite{scholz2003b}) on $\epsilon$~Indi system.}.

\begin{acknowledgements}
We thank our colleagues T.~Granzer (Astrophysikalisches Institut
Potsdam) and R.~Schwarz (Universit\"ats-Sternwarte G\"ottingen)
for their assistance in the reduction of DFOSC photometric data.
We have made use of the \ros\ Data Archive of the
Max-Planck-Institut f\"ur extraterrestrische Physik (MPE) at
Garching, Germany, data products from the SuperCOSMOS Sky Surveys
at the Wide-Field Astronomy Unit of the Institute for Astronomy,
University of Edinburgh and of the ESO Skycat Tool (version
2.5.3.). We have also used data products from the Two Micron All
Sky Survey, which is a joint project of the University of
Massachusetts and the Infrared Processing and Analysis
Center/California Institute of Technology, funded by the National
Aeronautics and Space Administration and the National Science
Foundation. This project is supported in part by the German BMBF
under DLR grant 50 OX 0201.

\end{acknowledgements}

\end{document}